\documentclass[a4paper]{article}
\usepackage{amssymb}
\usepackage{amsfonts}
\usepackage{graphicx}
\usepackage{amsmath}
\usepackage{float}
\newtheorem{theorem}{Theorem}[section]

\newtheorem{definition}{Definition}[section]

\begin{document}
\title{Quantum Experimental Data in Psychology and Economics}
\author{Diederik Aerts$^*$, Bart D'Hooghe\thanks{Leo Apostel Center for Interdisciplinary Studies, Brussels Free University,
Belgium. E-mails: diraerts@vub.ac.be, bdhooghe@vub.ac.be}\ and Emmanuel Haven\thanks{School of Management, University of Leicester, United Kingdom. E-mail: e.haven@le.ac.uk}}
\date{}
\maketitle

\begin{abstract}
We prove a theorem which shows that a collection of experimental data of
probabilistic weights related to decisions with respect to situations and their disjunction
cannot be modeled within a classical probabilistic weight structure in case the experimental data contain the effect referred to as the `disjunction effect' in psychology. We identify different experimental situations in
psychology, more specifically in concept theory and in decision theory, and in economics (namely situations where Savage's Sure-Thing Principle is violated) where the disjunction effect appears and we point out
the common nature of the effect. We analyze how our theorem constitutes a
no-go theorem for classical probabilistic weight structures for common experimental
data when the disjunction effect is affecting the values of these data.
We put forward a simple geometric criterion
that reveals the non classicality of the considered probabilistic weights and we illustrate our geometrical
criterion by means of 
experimentally measured membership weights of items with respect to pairs of concepts and their disjunctions. The violation of the classical probabilistic weight structure is very analogous
to the violation of the well-known Bell inequalities studied in quantum
mechanics. The no-go theorem we prove in the present article  with respect to the collection of experimental data we consider has a status analogous to the well known no-go theorems for hidden variable theories in quantum mechanics with respect to experimental data obtained in quantum laboratories.
For this reason our analysis puts forward a strong argument in favor of the validity of using a quantum formalism for modeling the considered psychological experimental data as considered in this paper.
\end{abstract}

\section{Introduction}
There exists an intensive ongoing research activity focusing on the use of the mathematical formalism of quantum mechanics to model situations in cognition and economics \cite{bruzagabora2009,bruzalawlessvanrijsbergensofte2007,bruzalawlessvanrijsbergensoftecoeckeclark2008,bruzasoftevanrijsbergenklusch2009}. Our group at the Leo Apostel Center in Brussels has played a role in the initiation of this research domain \cite{aertsaerts1994,aertsbroekaertsmets1999a,aertsbroekaertsmets1999b,aertsaertsbroekaertgabora2000,gaboraaerts2002,aertsczachor2004,aertsgabora2005a,aertsgabora2005b,broekaertaertsdhooghe2006,aerts2007}, and is still actively engaged in it \cite{aerts2009a,aerts2009b,aertsaertsgabora2009,aertsdhooghe2009,aerts2010a,aerts2010b,aertsbroekaertgabora2010,aertsdhooghe2010}.

In the present article we make use of insights and techniques developed in the foundations of quantum mechanics to investigate whether a specific collection of experimental data can be modeled by means of a classical theory, or whether a more general theory is needed, eventually a quantum theory. For decades intensive research has been conducted focussing on the very question, since physicists wanted to know whether quantum mechanics itself could be substituted by a classical theory. This body of research is traditionally referred to as `the hidden variable problem of quantum mechanics', because indeed such a classical theory giving rise to the same predictions as quantum mechanics would be a theory containing `hidden variables', to account for classical determinism on a hidden not necessarily manifest level. The presence of quantum-type probabilities would occur as a consequence of the lack of knowledge of these hidden variables on the manifest level. Physicists had already encountered such a situation before, namely classical statistical mechanics is a hidden variable theory for thermodynamics, i.e. the positions and velocities of the molecules of a given substance are hidden variables when the thermodynamic description level of the substance is the manifest level \cite{vonneumann1932,einsteinpodolskyrosen1935,bohr1935,gleason1957,jauchpiron1963,bell1964,bohmbub1966,bell1966,kochenspecker1967,jauchpiron1968,gudder1968,bohmbub1968,clauserhorneshimonyholt1969,gudder1970,clauserhorne1974,accardifedullo1982,aspectgrangierroger1982,aspectdalibardroger1982,accardi1984,aerts1985,aerts1986,aerts1987,redhead1987,pitowsky1989}. John von Neumann proved the first no-go theorem for the existence of hidden variables for quantum mechanics \cite{vonneumann1932}, and this was followed by the famous Einstein-Podolsky-Rosen paradox proposal \cite{einsteinpodolskyrosen1935}. Critical investigations with respect to both von Neumann's no-go theorem and the EPR paradox were performed by Bell \cite{bell1964,bell1966}. Then followed the proposal of an effective hidden variable theory, nowadays called `Bohm's theory', \cite{bohmbub1966}, elaborations of von Neumann's theorem, i.e. further investigations from a structural perspective \cite{gleason1957,jauchpiron1963,kochenspecker1967,gudder1970,accardifedullo1982,accardi1984,aerts1985,aerts1986,aerts1987,pitowsky1989}, and extensive discussions about several aspects of the problem \cite{jauchpiron1968,gudder1968,bohmbub1968,redhead1987}. In the seventies the experimentalist became interested, and this lead to new developments, e.g. the sharpening of notions such as locality, separability, etc.... But most of all, quantum mechanics was now confirmed as a profoundly reliable physical theory, even when scrutinized under all types of aspects where failure could be expected in a plausible way \cite{clauserhorneshimonyholt1969,clauserhorne1974,aspectgrangierroger1982,aspectdalibardroger1982}. In the eighties, it was shown, step by step, that by focusing on the mathematical structure of the probability model used to model experimental data, it was possible to distinguish between data that are quantum (more correctly `non-classical', in the sense of not allowing a modeling within a classical Kolmogorovian probability model \cite{kolmogorov1956}), and data that are classical (hence can be modeled within a Kolmogorovian probability model) \cite{accardifedullo1982,accardi1984,aerts1985,aerts1986,aerts1987,pitowsky1989}. One of the aspects of this hidden variable research, which from the foundations of quantum mechanics point of view is definitely of more universal importance and value, is that the results with respect to the characterization of a set of experimental data, i.e. whether these data can be modeled within a classical theory or not, do not depend on whether these data are obtained from measurements in a physics laboratory. For sets of data whether obtained from experiments in psychology or economics (or in any other domain of science) the same analysis can be made, and the same techniques of characterization of these data can be employed.

We have already investigated in this way data that were gathered by experiments measuring membership weights of an item with respect to two concepts and the conjunction of these two concepts \cite{hampton1988a}. These experimental data provide experimental evidence for a quantum structure in cognition \cite{aertsaertsgabora2009}. The deviation of what a classical probability theory would provide in modeling these experimental data was called `overextension' in concept research circles \cite{hampton1988a}. Many experiments by different concept researchers have measured the presence of `overextension' for the conjunction of concepts \cite{hampton1987,hampton1991,stormsdeboeckvanmechelengeeraerts1993,hampton1996,hampton1997,stormsdeboeckhamptonvanmechelen1999}, such that the `deviation from classicality' is experimentally well documented and abundant. There is a correspondence between `overextension' for typicality and membership weight values for the conjunction of concept, and what in decision theory is referred to as `the conjunction fallacy' \cite{tverskykahneman1982,tverskykahneman1983}. In the present article we want to concentrate on the `disjunction', and `how deviations from classicality appear when the disjunction is at play'. The experimental data that we consider as our elements of study are the results of measurements of membership weights of items with respect to pairs of concepts and their disjunction \cite{hampton1988b}. In an analogous way like conjunction deviations from classicality in concept theories relate to the conjunction fallacy in decision theory, there is the well studied disjunction effect in decision theory which corresponds to these disjunction deviations in concept theories \cite{carlsonyates1989,tverskyshafir1992,bar-hillelneter1993,croson1999,kuhbergerkomunskaperner2001,litaplin2002,vandijkzeelenberg2003,vandijkzeelenberg2006,bauerjohnson-laird2006,lambdinburdsal2007,bagassimacchi2007,hristovagrinberg2008}. In economics too, an effect similar to this disjunction effect was observed, more specifically in situations where Savage's Sure-Thing Principle \cite{savage1954}, a fundamental hypothesis of classical economic theory, is violated. We refer here to the Allais and Ellsberg paradoxes \cite{allais1953,ellsberg1961}.

Before putting forward a simple criterion and also a geometric interpretation of it in the next section, we would like to mention that the disjunction effect in decision theory has been modeled quantum mechanically by several authors \cite{busemeyerwangtownsend2006,pothosbusemeyer2009,khrennikovhaven2009}. The disjunction effect, as it appears for the membership weights of items with respect to pairs of concepts and their disjunction, was modeled explicitly by the quantum mechanical formalism in our Brussels group \cite{aerts2007,aerts2009a,aerts2010a}. The result of the present article, namely that the disjunction effect cannot be modeled classically, supports the quantum models that have been put forward for it.

\section{Classical and non classical membership weights for concepts}
The `disjunction' experiments we want to focus on in the present article were performed with the aim of measuring deviations for membership weights of items with respect to concepts from how one would expect such membership weights to behave classically \cite{hampton1988b}. For example, the concepts {\it Home Furnishings} and {\it Furniture} and their disjunction `{\it Home Furnishings or Furniture}' are considered. With respect to this pair, the item {\it Ashtray} is considered. Subjects rated the membership weight of {\it Ashtray} for the concept {\it Home Furnishings} as 0.7 and the membership weight of the item {\it Ashtray} for the concept {\it Furniture} as 0.3. However, the membership weight of {\it Ashtray} with respect to the disjunction `{\it Home Furnishings or Furniture}' was rated as only 0.25, i.e. less than either one of the weights assigned for both concepts separately. This means that subjects found {\it Ashtray} to be `less strongly a member of the disjunction `{\it Home Furnishings or Furniture}' than they found it to be a member of the concept {\it Home Furnishings} alone or a member of the concept {\it Furniture} alone'. If one thinks intuitively about the `logical' meaning of a disjunction, then this is an unexpected result. Indeed, if somebody finds that {\it Ashtray} belongs to {\it Home Furnishings}, they would be expected to also believe that {\it Ashtray} belongs to `{\it Home Furnishings or Furniture}'. The same holds for {\it Ashtray} and {\it Furniture}. Hampton called this deviation (this relative to what one would expect according to a standard classical interpretation of the disjunction) `underextension' \cite{hampton1988b}. 

A typical experiment testing the effect described above proceeds as follows. The tested subjects are asked to choose a number from the following set: $\{-3,-2, -1,0,$ $+1,+2,+3\}$, where the positive numbers +1, +2 or +3 mean that they consider `the item to be a member of the concept' and the typicality of the membership increases with an increasing number. Hence +3 means that the subject who attributes this number considers the item to be a very typical member, and +1 means that he or she considers the item to be a not so typical member. The negative numbers indicate non-membership, again in increasing order, i.e. -3 indicates strong non-membership, and -1 represents weak non-membership. Choosing 0 indicates the subject is indecisive about the membership or non-membership of the item. Subjects were asked to repeat the procedure for all the items and concepts considered. Membership weights were then calculated by dividing the number of positive ratings by the number of non-zero ratings.

Considering again the case of {\it Ashtray} as an item and its membership with respect to the concepts {\it Home Furnishings} and {\it Furniture} and their disjunction. As the experiments are conceived, each individual subject will decide for {\it Ashtray} whether it is a member or not a member of respectively {\it Home Furnishings}, {\it Furniture} and `{\it Home Furnishings or Furniture}'. Suppose that there are $n$ subjects participating in the experiment. There is a way to express what we mean intuitively by `classical behavior'. Indeed, what we would `not' like to happen is that a subject, decides {\it Ashtray} to be a member of {\it Home Furnishings}, but not a member of {\it Home Furnishings or Furniture}. If a subject would make such type of decision, then this would be in direct conflict with the meaning of the disjunction. However, in the case of {\it Ashtray}, since $0.7 \times n$ subjects have decided that {\it Ashtray} is a member of {\it Home Furnishings} and only $0.25 \times n$ subjects have decided it to be a member of `{\it Home Furnishings or Furniture}', this means that at least $0.55 \times n$ subjects have taken this decision in direct conflict with the meaning of the disjunction. In case $n=100$, this means $55$ subjects (more than half) have done so.

Suppose we introduce the following notation, and indicate with $A_1$ the first considered concept, hence {\it Home Furnishings}, and with $\mu(A_1)$ the membership weight of item $X$, hence {\it Ashtray}, with respect to $A_1$. This means that for our example we have $\mu(A_1)=0.7$. With $A_2$ we denote the second considered concept, hence {\it Furniture}, and with $\mu(A_2)$ the membership weight of item $X$, hence {\it Ashtray}, with respect to $A_2$. This means that for our example we have $\mu(A_2)=0.3$. With `$A_1$ or $A_2$' we denote the disjunction of both concepts $A_1$ and $A_2$, hence `{\it Home Furnishings or Furniture}', and with $\mu(A_1\ {\rm or}\ A_2)$ the membership weight of item $X$, hence {\it Ashtray}, with respect to `$A_1$ or $A_2$'.

We can easily see that the non classical effect we analyzed above cannot happen in case the following two inequalities are satisfied 
\begin{eqnarray} \label{A1leA1orA2}
&&\mu(A_1) \le \mu(A_1\ {\rm or}\ A_2) \\ \label{A2leA1orA2}
&&\mu(A_2) \le \mu(A_1\ {\rm or}\ A_2)
\end{eqnarray} 
and we observe indeed that both inequalities are violated for our example of {\it Ashtray} with respect to {\it Home Furnishings} and {\it Furniture}.

There is another issue which we do not want to happen, and this one is somewhat more subtle. To illustrate it, we consider another example of the experiments, namely the item {\it Olive}, with respect to the pair of concepts {\it Fruits} and {\it Vegetables} and their disjunction `{\it Fruits or Vegetables}'. The respective membership weights were measured to be $\mu(A_1)=0.5$, $\mu(A_2)=0.1$ and $\mu(A_1\ {\rm or}\ A_2)=0.8$. Obviously inequalities (\ref{A1leA1orA2}) and (\ref{A2leA1orA2}) are both satisfied for this example. Let us suppose again that there are $n$ subjects participating in the experiment. Then $0.5 \times n$ subjects have decided that {\it Olive} is a member of {\it Fruits}, and $0.1 \times n$ subjects have decided that {\it Olive} is a member of {\it Vegetables}, while $0.8 \times n$ subjects have decided that {\it Olive} is a member of `{\it Fruits or Vegetables}'. However, at maximum $0.5 \times n + 0.1 \times n = 0.6 \times n$ subjects have decided that {\it Olive} is a member of {\it Fruits} `or' is a member of {\it Vegetables}. This means that a minimum of $0.4 \times n$ subjects have decided that {\it Olive} is neither a member of {\it Fruits} nor a member of {\it Vegetables}. But $0.8 \times n$ subjects have decided that {\it Olive} is a member of `{\it Fruits or Vegetables}'. This means that a minimum of $0.2 \times n$ subjects have decided that {\it Olive} `is not' a member of {\it Fruits}, and also `is not' a member of {\it Vegetables}, but `is' a member of `{\it Fruits or Vegetables}'. The decision made by these $0.2 \times n$ subjects, hence 20 in case $n=100$, goes directly against the meaning of the disjunction. An item becoming a member of the disjunction while it is not a member of both pairs is completely non classical. We can easily see that this second type of non classicality cannot happen in case the following inequality is valid
\begin{equation} \label{A1orA2leA1plusA2}
\mu(A_1\ {\rm or}\ A_2) \le \mu(A_1)+\mu(A_2)
\end{equation}
and indeed this inequality is violated by the example of {\it Olive} with respect to {\it Fruits} and {\it Vegetables}. One of the authors has derived in earlier work the three inequalities (\ref{A1leA1orA2}), (\ref{A2leA1orA2}) and (\ref{A1orA2leA1plusA2}) as a consequence of a different type of requirement, namely the requirement that the membership weights are in their most general form representations of mathematical normed measures (see section 1.4, Theorem 4 and Appendix B of \cite{aerts2009a}, and Theorem 4.1 of the present article). The items that deviated from classicality by violating one or both of the inequalities (\ref{A1leA1orA2}) and (\ref{A2leA1orA2}) were called $\Delta$-type non classical items. The items that deviated from classicality by violating inequality (\ref{A1orA2leA1plusA2}) were called $k$-type non classical items. An explicit quantum model was constructed for both types of non classical items \cite{aerts2009a}. In the present paper we consider the general situation of $n$ concepts and disjunctions of pairs of these $n$ concepts, and we introduce the corresponding definition for classicality. We will additionally derive a simple geometrical criterion to verify whether the membership weights of an item with respect to a set of concepts and disjunctions of pairs of them can be modeled classically or not. Table 1 represents the items and pairs of concepts tested by Hampton \cite{hampton1988b} which we will use as experimental data to illustrate the analysis put forward in the present article.

We consider $n$ concepts $A_{1},A_{2},\ldots ,A_{n}$ and membership weights $\mu
(A_{i})$ of an item $X$ with respect to each concept $A_{i}$, and also membership weights $\mu (A_{i}\ \mathrm{or}\
A_{j})$ of this item $X$ with respect to the disjunction of concepts $A_{i}$ and $A_{j}$. It is not necessary that membership weights of the item $X$ are determined with respect to each one of the possible
pairs of concepts. Hence, to describe this situation formally, we consider a
set $S$ of pairs of indices $S\subseteq \{(i,j)\ |\ i<j;i,j=1,2,\ldots ,n\}$
corresponding to those pairs of concepts for which the membership weights of the item $X$ have been
measured with respect to the disjunction of these pairs. Hence, the following set of membership weights have been experimentally determined 
\begin{equation}
p_{i}=\mu (A_{i})\quad i=1,2,\ldots ,n;\quad p_{i\vee j}=\mu (A_{i}\ \mathrm{%
or}\ A_{j})\quad (i,j)\in S  \label{weightsp}
\end{equation}%
\begin{definition}[Classical Disjunction Data]
We say that the set of membership weights of an item $X$ with respect to concepts is a `classical disjunctive set of membership
weights' if it has a normed measure representation. Hence if there exists a
normed measure space $(\Omega ,\sigma (\Omega ),P)$ with $%
E_{A_{1}},E_{A_{2}},\ldots ,E_{A_{n}}\in \sigma (\Omega )$ elements of the
event algebra, such that 
\begin{equation}
p_{i}=P(E_{A_{i}})\quad i=1,2,\ldots ,n;\quad p_{i\vee j}=P(E_{A_{i}}\cup
E_{A_{j}})\quad (i,j)\in S  \label{weightsalgebra}
\end{equation}
\end{definition}
A normed measure $P$ is a function defined on a $\sigma $-algebra $\sigma
(\Omega )$ over a set $\Omega $ and taking values in the interval $[0,1]$
such that the following properties are satisfied: (i) The empty set has
measure zero, i.e. $P(\emptyset )=0$; (ii) Countable additivity or $\sigma $%
-additivity: if $E_{1}$, $E_{2}$, $E_{3}$, $\dots $ is a countable sequence
of pairwise disjoint sets in $\sigma (\Omega )$, the measure of the union of
all the $E_{i}$ is equal to the sum of the measures of each $E_{i}$, i.e. $%
P(\bigcup_{i=1}^{\infty }E_{i})=\sum_{i=1}^{\infty }P(E_{i})$; (iii) The
total measure is one, i.e. $P(\Omega )=1$. The triple $(\Omega ,\sigma
(\Omega ),P)$ is called a normed measure space, and the members of $\sigma
(\Omega )$ are called measurable sets. A $\sigma $-algebra over a set $%
\Omega $ is a

\begin{table}
\caption{The eight pairs of concepts and items of experiment 2 in \cite{hampton1988b}. $\mu(A_1)$, $\mu(A_2)$ and $\mu(A_1{\rm or}A_2)$ are respectively the measured membership weights of each item with respect to the concepts $A_1$, $A_2$ and their disjunction $A_1$or$A_2$. The non classical items are labeled by $q$ and the classical items by $c$.}
\begin{center}
\footnotesize
\begin{tabular}{|lllll|lllll|}
\hline 
\multicolumn{1}{|l}{} & \multicolumn{1}{l}{} & \multicolumn{1}{l}{$\mu(A_1)$} & \multicolumn{1}{l}{$\mu(A_2)$} & \multicolumn{1}{l|}{$\mu(A_1{\rm or}A_2)$} & \multicolumn{1}{l}{} & \multicolumn{1}{l}{} &\multicolumn{1}{l}{$\mu(A_1)$} & \multicolumn{1}{l}{$\mu(A_2)$} & \multicolumn{1}{l|}{$\mu(A_1{\rm or}A_2)$} \\
\hline
\multicolumn{5}{|l|}{\it $A_1$=Home Furnishing, $A_2$=Furniture} & \multicolumn{5}{l|}{\it $A_1$=Spices, $A_2$=Herbs} \\
\hline
{\it Mantelpiece} & $q$ & 0.8 & 0.4 & 0.75 & {\it Molasses} & $c$ & 0.4 & 0.05 & 0.425 \\
{\it Window Seat } & $q$ & 0.9 & 0.9 & 0.8 & {\it Salt} & $q$ & 0.75 & 0.1 & 0.6 \\
{\it Painting } & $q$ & 0.9 & 0.5 & 0.85 & {\it Peppermint} & $c$ & 0.45 & 0.6 & 0.6 \\
{\it Light Fixture} & $q$ & 0.8 & 0.4 & 0.775 & {\it Curry} & $q$ & 0.9 & 0.4 & 0.75 \\
{\it Kitchen Count} & $q$ & 0.8 & 0.55 & 0.625 &   {\it Oregano} & $q$ & 0.7 & 1 & 0.875 \\ 
{\it Bath Tub} & $c$ & 0.5 & 0.7 & 0.75 & {\it MSG} & $q$ & 0.15 & 0.1 & 0.425 \\
{\it Desk Chair} & $c$ & 0.1 & 0.3 & 0.35 & {\it Chili Pepper} & $q$ & 1 & 0.6 & 0.95 \\
{\it Shelves} & $c$ & 1 & 0.4 & 1 & {\it Mustard} & $q$ & 1 & 0.8 & 0.85  \\
{\it Rug} & $c$ & 0.9 & 0.6 & 0.95 & {\it Mint} & $c$ & 1 & 0.8 & 0.925 \\
{\it Bed} & $c$ & 1 & 1 & 1 & {\it Cinnamon} & $c$ & 1 & 0.4 & 1 \\
{\it Wall-Hangings} & $c$ & 0.9 & 0.4 & 0.95 & {\it Parsley} & $c$ & 0.5 & 0.9 & 0.95 \\
{\it Space Rack} & $q$ & 0.7 & 0.5 & 0.65 & {\it Saccharin} & $q$ & 0.1 & 0.01 & 0.15 \\
{\it Ashtray} & $q$ & 0.7 & 0.3 & 0.25 & {\it Poppyseeds} & $c$ & 0.4 & 0.4 & 0.4 \\
{\it Bar} & $q$ & 0.35 & 0.6 & 0.55 & {\it Pepper} & $c$ & 0.9 & 0.6 & 0.95 \\
{\it Lamp} & $q$ & 1 & 0.7 & 0.9 & {\it Turmeric} & $q$ & 0.7 & 0.45 & 0.675 \\
{\it Wall Mirror} & $q$ & 1 & 0.6 & 0.95 & {\it Sugar} & $q$ & 0 & 0 & 0.2 \\
{\it Door Bell} & $c$ & 0.5 & 0.1 & 0.55 & {\it Vinegar} & $q$ & 0.1 & 0.01 & 0.35 \\
{\it Hammock} & $q$ & 0.2 & 0.5 & 0.35 & {\it Sesame Seeds} & $c$ & 0.35 & 0.4 & 0.625 \\
{\it Desk} & $c$ & 1 & 1 & 1 & {\it Lemon Juice} & $q$ & 0.1 & 0.01 & 0.15 \\
{\it Refrigerator} & $q$ & 0.9 & 0.7 & 0.575 & {\it Chocolate} & $c$ & 0 & 0 & 0 \\
{\it Park Bench} & $q$ & 0 & 0.3 & 0.05 & {\it Horseradish} & $q$ & 0.2 & 0.4 & 0.7 \\
{\it Waste Paper Basket} & $q$ & 1 & 0.5 & 0.6 & {\it Vanilla} & $q$ & 0.6 & 0 & 0.275 \\
{\it Sculpture} & $c$ & 0.8 & 0.4 & 0.8 & {\it Chires} & $q$ & 0.6 & 1 & 0.95 \\
{\it Sink Unit} & $q$ & 0.9 & 0.6 & 0.6 & {\it Root Ginger} & $q$ & 0.7 & 0.15 & 0.675 \\
\hline
\multicolumn{5}{|l|}{\it $A$=Hobbies, $B$=Games} & \multicolumn{5}{l|}{\it $A$=Instruments, $B$=Tools} \\
\hline
{\it Gardening} & $c$ & 1 & 0 & 1 & {\it Broom} & $q$ & 0.1 & 0.7 & 0.6 \\
{\it Theatre-Going} & $c$ & 1 & 0 & 1 & {\it Magnetic Compass} & $c$ & 0.9 & 0.5 & 1 \\
{\it Archery} & $q$ & 1 & 0.9 & 0.95 & {\it Tuning Fork} & $c$ & 0.9 & 0.6 & 1 \\
{\it Monopoly} & $c$ & 0.7 & 1 & 1 & {\it Pen-Knife} & $q$ & 0.65 & 1 & 0.95 \\
{\it Tennis} & $c$ & 1 & 1 & 1 & {\it Rubber Band} & $q$ & 0.25 & 0.5 & 0.25 \\
{\it Bowling} & $c$ & 1 & 1 & 1 & {\it Stapler} & $c$ & 0.85 & 0.8 & 0.85 \\
{\it Fishing} & $c$ & 1 & 0.6 & 1 & {\it Skate Board} & $q$ & 0.1 & 0 & 0 \\
{\it Washing Dishes} & $q$ & 0.1 & 0 & 0.15 & {\it Scissors} & $q$ & 0.85 & 1 & 0.9 \\
{\it Eating Ice-Cream Cones} & $q$ & 0.2 & 0 & 0.1 & {\it Pencil Eraser} & $q$ & 0.4 & 0.7 & 0.45  \\
{\it Camping} & $q$ & 1 & 0.1 & 0.9 & {\it Tin Opener} & $c$ & 0.9 & 0.9 & 0.95  \\
{\it Skating} & $q$ & 1 & 0.6 & 0.95 & {\it Bicycle Pump} & $q$ & 1 & 0.9 & 0.7 \\
{\it Judo} & $q$ & 1 & 0.7 & 0.8 & {\it Scalpel} & $q$ & 0.8 & 1 & 0.925 \\
{\it Guitar Playing} & $c$ & 1 & 0 & 1 & {\it Computer} & $q$ & 0.6 & 0.8 & 0.6 \\
{\it Autograph Hunting} & $q$ & 1 & 0.2 & 0.9 & {\it Paper Clip} & $q$ & 0.3 & 0.7 & 0.6 \\
{\it Discus Throwing} & $q$ & 1 & 0.75 & 0.7 & {\it Paint Brush} & $c$ & 0.65 & 0.9 & 0.95 \\
{\it Jogging} & $q$ & 1 & 0.4 & 0.9 & {\it Step Ladder} & $q$ & 0.2 & 0.9 & 0.85 \\
{\it Keep Fit} & $q$ & 1 & 0.3 & 0.95 & {\it Door Key} & $q$ & 0.3 & 0.1 & 0.95 \\
{\it Noughts} & $q$ & 0.5 & 1 & 0.9 & {\it Measuring Calipers} & $q$ & 0.9 & 1 & 0.9 \\
{\it Karate} & $q$ & 1 & 0.7 & 0.8 & {\it Toothbrush} & $c$ & 0.4 & 0.4 & 0.5 \\
{\it Bridge} & $c$ & 1 & 1 & 1 & {\it Sellotape} & $q$ & 0.1 & 0.2 & 0.325 \\
{\it Rock Climbing} & $q$ & 1 & 0.2 & 0.95 & {\it Goggles} & $q$ & 0.2 & 0.3 & 0.15 \\
{\it Beer Drinking} & $q$ & 0.8 & 0.2 & 0.575 & {\it Spoon} & $q$ & 0.65 & 0.9 & 0.7 \\
{\it Stamp Collecting} & $c$ & 1 & 0.1 & 1 & {\it Pliers} & $c$ & 0.8 & 1 & 1 \\
{\it Wrestling} & $q$ & 0.9 & 0.6 & 0.625 & {\it Meat Thermometer} & $c$ & 0.75 & 0.8 & 0.9 \\
\hline
\end{tabular}
\end{center}
\end{table}

\begin{center}
\footnotesize
\begin{tabular}{|lllll|lllll|}
\hline 
\multicolumn{1}{|l}{} & \multicolumn{1}{l}{} & \multicolumn{1}{l}{$\mu(A_1)$} & \multicolumn{1}{l}{$\mu(A_2)$} & \multicolumn{1}{l|}{$\mu(A_1{\rm or}A_2)$} & \multicolumn{1}{l}{} & \multicolumn{1}{l}{} &\multicolumn{1}{l}{$\mu(A_1)$} & \multicolumn{1}{l}{$\mu(A_2)$} & \multicolumn{1}{l|}{$\mu(A_1{\rm or}A_2)$} \\
\hline
\multicolumn{5}{|l|}{\it $A_1$=Pets, $A_2$=Farmyard Animals} & \multicolumn{5}{l|}{\it $A_1$=Fruits, $A_2$=Vegetables} \\
\hline
{\it Goldfish} & $q$ & 1 & 0 & 0.95 & {\it Apple} & $c$ & 1 & 0 & 1 \\
{\it Robin} & $c$ & 0.1 & 0.1 & 0.1 & {\it Parsley} & $q$ & 0 & 0.2 & 0.45 \\
{\it Blue-Tit} & $c$ & 0.1 & 0.1 & 0.1 & {\it Olive} & $q$ & 0.5 & 0.1 & 0.8 \\
{\it Collie Dog} & $c$ & 1 & 0.7 & 1 & {\it Chili Pepper} & $c$ & 0.05 & 0.5 & 0.5 \\
{\it Camel} & $q$ & 0.4 & 0 & 0.1 &   {\it Broccoli} & $q$ & 0 & 0.8 & 1 \\ 
{\it Squirrel} & $q$ & 0.2 & 0.1 & 0.1 & {\it Root Ginger} & $q$ & 0 & 0.3 & 0.55 \\
{\it Guide Dog for the Blind} & $q$ & 0.7 & 0 & 0.9 & {\it Pumpkin} & $c$ & 0.7 & 0.8 & 0.925 \\
{\it Spider} & $c$ & 0.5 & 0.35 & 0.55 & {\it Raisin} & $q$ & 1 & 0 & 0.9  \\
{\it Homing Pig} & $q$ & 0.9 & 0.1 & 0.8 & {\it Acorn} & $q$ & 0.35 & 0 & 0.4 \\
{\it Monkey} & $q$ & 0.5 & 0 & 0.25 & {\it Mustard} & $q$ & 0 & 0.2 & 0.175 \\
{\it Circus Horse} & $q$ & 0.4 & 0 & 0.3 & {\it Rice} & $q$ & 0 & 0.4 & 0.325 \\
{\it Prize Bull} & $q$ & 0.1 & 1 & 0.9 & {\it Tomato} & $c$ & 0.7 & 0.7 & 1 \\
{\it Rat} & $q$ & 0.5 & 0.7 & 0.4 & {\it Coconut} & $q$ & 0.7 & 0 & 1 \\
{\it Badger} & $q$ & 0 & 0.25 & 0.1 & {\it Mushroom} & $q$ & 0 & 0.5 & 0.9 \\
{\it Siamese Cat} & $q$ & 1 & 0.1 & 0.95 & {\it Wheat} & $q$ & 0 & 0.1 & 0.2 \\
{\it Race Horse} & $c$ & 0.6 & 0.25 & 0.65 & {\it Green Pepper} & $c$ & 0.3 & 0.6 & 0.8 \\
{\it Fox} & $q$ & 0.1 & 0.3 & 0.2 & {\it Watercress} & $q$ & 0 & 0.6 & 0.8 \\
{\it Donkey} & $q$ & 0.5 & 0.9 & 0.7 & {\it Peanut} & $c$ & 0.3 & 0.1 & 0.4 \\
{\it Field Mouse} & $q$ & 0.1 & 0.7 & 0.4 & {\it Black Pepper} & $c$ & 0.15 & 0.2 & 0.225 \\
{\it Ginger Tom-Cat} & $q$ & 1 & 0.8 & 0.95 & {\it Garlic} & $q$ & 0.1 & 0.2 & 0.5 \\
{\it Husky in Sledream} & $q$ & 0.4 & 0 & 0.425 & {\it Yam} & $c$ & 0.45 & 0.65 & 0.85 \\
{\it Cart Horse} & $q$ & 0.4 & 1 & 0.85 & {\it Elderberry} & $q$ & 1 & 0 & 0.8 \\
{\it Chicken} & $q$ & 0.3 & 1 & 0.95 & {\it Almond} & $q$ & 0.2 & 0.1 & 0.425 \\
{\it Doberman Guard Dog} & $q$ & 0.6 & 0.85 & 0.8 & {\it Lentils} & $q$ & 0 & 0.6 & 0.525 \\
\hline
\multicolumn{5}{|l|}{\it $A$=Sportswear, $B$=Sports Equipment} & \multicolumn{5}{l|}{\it $A$=Household Appliances, $B$=Kitchen Utensils} \\
\hline
{\it American Foot} & $c$ & 1 & 1 & 1 & {\it Fork} & $q$ & 0.7 & 1 & 0.95 \\
{\it Referee's Whistle} & $q$ & 0.6 & 0.2 & 0.45 & {\it Apron} & $c$ & 0.3 & 0.4 & 0.5 \\
{\it Circus Clowns} & $q$ & 0 & 0 & 0.1 & {\it Hat Stand} & $q$ & 0.45 & 0 & 0.3 \\
{\it Backpack} & $c$ & 0.6 & 0.5 & 0.6 & {\it Freezer} & $q$ & 1 & 0.6 & 0.95 \\
{\it Diving Mask} & $q$ & 1 & 1 & 0.95 & {\it Extractor Fan} & $q$ & 1 & 0.4 & 0.9 \\
{\it Frisbee} & $q$ & 0.3 & 1 & 0.85 & {\it Cake Tin} & $c$ & 0.4 & 0.7 & 0.95 \\
{\it Sunglasses} & $q$ & 0.4 & 0.2 & 0.1 & {\it Carving Knife} & $c$ & 0.7 & 1 & 1 \\
{\it Suntan Lotion} & $q$ & 0 & 0 & 0.1 & {\it Cooking Stove} & $c$ & 1 & 0.5 & 1 \\
{\it Gymnasium} & $q$ & 0 & 0.9 & 0.825 & {\it Iron} & $q$ & 1 & 0.3 & 0.95  \\
{\it Motorcycle Helmet} & $q$ & 0.7 & 0.9 & 0.75 & {\it Food Processor} & $c$ & 1 & 1 & 1  \\
{\it Rubber Flipper} & $c$ & 1 & 1 & 1 & {\it Chopping Board} & $q$ & 0.45 & 1 & 0.95 \\
{\it Wrist Sweat} & $q$ & 1 & 1 & 0.95 & {\it Television} & $q$ & 0.95 & 0 & 0.85 \\
{\it Golf Ball} & $c$ & 0.1 & 1 & 1 & {\it Vacuum Cleaner} & $c$ & 1 & 0 & 1 \\
{\it Cheerleaders} & $c$ & 0.3 & 0.4 & 0.45 & {\it Rubbish Bin} & $c$ & 0.5 & 0.5 & 0.8 \\
{\it Linesman's Flag} & $q$ & 0.1 & 1 & 0.75 & {\it Vegetable Rack} & $c$ & 0.4 & 0.4 & 0.7 \\
{\it Underwater} & $q$ & 1 & 0.65 & 0.6 & {\it Broom} & $c$ & 0.55 & 0.4 & 0.625 \\
{\it Baseball Bat} & $c$ & 0.2 & 1 & 1 & {\it Rolling Pin} & $c$ & 0.45 & 1 & 1 \\
{\it Bathing Costume} & $q$ & 1 & 0.8 & 0.8 & {\it Table Mat} & $q$ & 0.25 & 0.4 & 0.325 \\
{\it Sailing Life Jacket} & $c$ & 1 & 0.8 & 1 & {\it Whisk} & $c$ & 1 & 1 & 1 \\
{\it Ballet Shoes} & $q$ & 0.7 & 0.6 & 0.6 & {\it Blender} & $c$ & 1 & 1 & 1 \\
{\it Hoola Hoop} & $q$ & 0.1 & 0.6 & 0.5 & {\it Electric Toothbrush} & $q$ & 0.8 & 0 & 0.55 \\
{\it Running Shoes} & $c$ & 1 & 1 & 1 & {\it Frying Pan} & $q$ & 0.7 & 1 & 0.95 \\
{\it Cricket Pitch} & $q$ & 0 & 0.5 & 0.525 & {\it Toaster} & $c$ & 1 & 1 & 1 \\
{\it Tennis Racket} & $c$ & 0.2 & 1 & 1 & {\it Spatula} & $c$ & 0.55 & 0.9 & 0.95 \\
\hline
\end{tabular}
\end{center}
\noindent
nonempty collection $\sigma (\Omega )$ of subsets of $\Omega $
that is closed under complementation and countable unions of its members.
Measure spaces are the most general structures devised by mathematicians and
physicists to represent weights.

\section{Geometrical characterization of membership weights}

We now develop the geometric language that makes it possible to verify the
existence of a normed measure representation for a set of weights. For this purpose we introduce the `classical disjunction polytope' $d_{c}\left( n,S\right)$ in the following way. We
construct an $n+\left\vert S\right\vert $ dimensional `classical disjunction
vector' 
\begin{equation*}
\overrightarrow{p}=\left( p_{1},p_{2},\ldots ,p_{n},\ldots ,p_{i\vee
j},\ldots \right)
\end{equation*}%
where $|S|$ is the cardinality of $S$. We consider the linear space $R\left(
n,S\right) \cong \mathbb{R}^{n+\left\vert S\right\vert }$ consisting of all
real vectors of this type. Next, let $\varepsilon \in \left\{ 0,1\right\}
^{n}$ be an arbitrary $n$-dimensional vector consisting of $0$ and $1$'s.
For each $\varepsilon $ we construct the classical disjunction vector $%
\overrightarrow{v}^{\varepsilon }\in R\left( n,S\right) $ by putting:%
\begin{equation*}
\begin{array}{cc}
v_{i}^{\varepsilon }=\varepsilon _{i} & i=1,\ldots ,n \\ 
v_{ij}^{\varepsilon }=\max \left( \varepsilon _{i},\varepsilon _{j}\right)
=\varepsilon _{i}+\varepsilon _{j}-\varepsilon _{i}\varepsilon _{j} & \left(
i,j\right) \in S%
\end{array}%
\end{equation*}%
The set of convex linear combinations of $\overrightarrow{v}^{\varepsilon }$
we call the `classical disjunction polytope' $d_{c}\left( n,S\right) :$%
\begin{equation*}
d_{c}\left( n,S\right) =\left\{ \overrightarrow{w}\in R\left( n,S\right)
\mid \overrightarrow{w}=\underset{\varepsilon \in \left\{ 0,1\right\} ^{n}}{%
\sum }\lambda _{\varepsilon }\overrightarrow{v}^{\varepsilon };\lambda
_{\varepsilon }\geq 0;\underset{\varepsilon \in \left\{ 0,1\right\} ^{n}}{%
\sum }\lambda _{\varepsilon }=1\right\}
\end{equation*}%
We prove now the following theorem
\begin{theorem}
The set of weights 
\begin{equation}
p_{i}=\mu (A_{i})\quad i=1,2,\ldots ,n;\quad p_{i\vee j}=\mu (A_{i}\ \mathrm{%
or}\ A_{j})\quad (i,j)\in S  \notag
\end{equation}%
admits a normed measure space, and hence is a classical disjunction set of
membership weights, if and only if its disjunction vector $\overrightarrow{p}
$ belongs to the classical disjunction polytope $d_{c}\left( n,S\right) $.
\end{theorem}
Proof: Suppose that (\ref{weightsp}) is a classical disjunction set of
weights, and hence we have a normed measure space $(\Omega ,\sigma (\Omega
),P)$ and events $E_{A_{i}}\in \sigma (\Omega )$ such that (\ref%
{weightsalgebra}) are satisfied. Let us show that in this case $%
\overrightarrow{p}\in d_{c}\left( n,S\right) $. For an arbitrary subset $%
X\subset \Omega $ we define $X^{1}=X$ and $X^{0}=\Omega \backslash X$.
Consider $\epsilon =(\epsilon _{1},\ldots ,\epsilon _{n})\in \{0,1\}^{n}$
and define $A(\epsilon )=\cap _{\epsilon }A_{i}^{\epsilon _{i}}$. Then we
have that $A(\epsilon )\cap A(\epsilon ^{\prime })=\emptyset $ for $\epsilon
\not=\epsilon ^{\prime }$, $\cup _{\epsilon }A(\epsilon )=\Omega $, and $%
\cup _{\epsilon ,\epsilon _{j}=1}A(\epsilon )=A_{j}$. We put now $\lambda
_{\epsilon }=P(A(\epsilon ))$. Then we have $\lambda _{\epsilon }\geq 0$ and 
$\sum_{\epsilon }\lambda _{\epsilon }=1$, and $p_{i}=P(A_{i})=\sum_{\epsilon
,\epsilon _{i}=1}\lambda _{\epsilon }=\sum_{\epsilon }\lambda _{\epsilon
}\epsilon _{i}$. We also have $p_{i\vee j}=P(A_{i}\cup A_{j})=\sum_{\epsilon
,\max \left( \epsilon _{i},\epsilon _{j}\right) =1}\lambda _{\epsilon
}=\sum_{\epsilon }\lambda _{\epsilon }\left( \epsilon _{i}+\epsilon
_{j}-\epsilon _{i}\epsilon _{j}\right) $. This means that $\overrightarrow{p}%
=\sum_{\epsilon }\lambda _{\epsilon }v^{\epsilon }$, which shows that $%
\overrightarrow{p}\in d_{c}\left( n,S\right) $. Conversely, suppose that $%
\overrightarrow{p}\in d_{c}\left( n,S\right) $. Then there exist numbers $%
\lambda _{\epsilon }\geq 0$ such that $\sum_{\epsilon }\lambda _{\epsilon }=1
$ and $\overrightarrow{p}=\sum_{\epsilon }\lambda _{\epsilon }v^{\epsilon }$%
. We define $\Omega =\{0,1\}^{n}$ and $\sigma (\Omega )$ the power set of $%
\Omega $. For $X\subset \Omega $ we define then $P(X)=\sum_{\epsilon \in
X}\lambda _{\epsilon }$. Then we choose $A_{i}=\{\epsilon ,\epsilon _{i}=1\}$
which gives that $P(A_{i})=\sum_{\epsilon }\lambda _{\epsilon }\epsilon
_{i}=\sum_{\epsilon }\lambda _{\epsilon }v_{i}^{\epsilon }=p_{i}$ and $%
P(A_{i}\cup A_{j})=\sum_{\epsilon }\lambda _{\epsilon }\left( \epsilon
_{i}+\epsilon _{j}-\epsilon _{i}\epsilon _{j}\right) =\sum_{\epsilon
}\lambda _{\epsilon }v_{ij}^{\epsilon }=p_{i\vee j}$. This shows that we
have a classical disjunction set of weights. {\bf QED}

\bigskip
\noindent
As one may notice, these results are very similar to those of Pitowsky for
classical conjunction polytopes $c\left( n,S\right)$ \cite{pitowsky1989}. However, the Pitowsky
correlation polytope and the classical disjunction polytope have different
sets of vertices. Furthermore, the interpretation of the $|S|$ components is completely different, namely representing conjunction data $p_{ij}$ and
disjunction data $p_{i\vee j}$ respectively. In general, the existence of a
classical disjunctive representation does not necessarily imply the
existence of a classical conjunctive representation, and vice versa.
Therefore, in order to fully grasp classicality by these geometric means,
the natural next step is to combine the theoretical results for conjunction
(Pitowsky) and disjunction polytopes (developed here and in \cite{aerts2009a}) by introducing a `classical
connective polytope'.

Again, let $\varepsilon \in \left\{ 0,1\right\} ^{n}$ be an arbitrary $n$%
-dimensional vector consisting of $0$ and $1$'s. For each $\varepsilon $ we
construct the classical connective vector $\overrightarrow{w}^{\varepsilon
}\in \mathbb{R}^{n+\left\vert S\right\vert +\left\vert S^{\prime
}\right\vert }$ by putting:%
\begin{equation*}
\begin{array}{cc}
w_{i}^{\varepsilon }=\varepsilon _{i} & i=1,\ldots ,n \\ 
w_{ij}^{\varepsilon }=\varepsilon _{i}\varepsilon _{j}=\min \left(
\varepsilon _{i},\varepsilon _{j}\right) & \left( i,j\right) \in S \\ 
w_{k\vee l}^{\varepsilon }=\varepsilon _{k}+\varepsilon _{l}-\varepsilon
_{k}\varepsilon _{l}=\max \left( \varepsilon _{k},\varepsilon _{l}\right) & 
\left( k,l\right) \in S^{\prime }%
\end{array}%
\end{equation*}%
The set of convex linear combinations of $\overrightarrow{w}^{\varepsilon }$
we call the `classical connective polytope' $k(n,S,S^{\prime })$: 
\begin{equation}
k(n,S,S^{\prime })=\{\overrightarrow{f}\in \mathbb{R}^{n+\left\vert
S\right\vert +\left\vert S^{\prime }\right\vert }\ |\ \overrightarrow{f}%
=\sum_{\epsilon \in \{0,1\}^{n}}\lambda _{\epsilon }\overrightarrow{w}%
^{\epsilon };\ \lambda _{\epsilon }\geq 0;\ \sum_{\epsilon \in
\{0,1\}^{n}}\lambda _{\epsilon }=1\}
\end{equation}

\begin{theorem}
The set of weights 
\begin{equation}
p_{i}=\mu (A_{i})\quad i=1,2,\ldots ,n;\quad p_{ij}=\mu (A_{i}\ \mathrm{and}%
\ A_{j})\quad (i,j)\in S;p_{i\vee j}=\mu (A_{i}\ \mathrm{or}\ A_{j})\quad
(i,j)\in S^{\prime }  \notag  \label{weights02}
\end{equation}%
admits a normed measure space, and hence is a classical set of membership
weights, if and only if its connective vector $\overrightarrow{p}$ belongs
to the classical connective polytope $k(n,S,S^{\prime }).$
\end{theorem}

Proof: Follows from the theorems for conjunction and disjunction classicality.

\section{A simple case: disjunction effect for 2 concepts}

One of the authors studied the disjunction effect for the case of two concepts and their disjunction \cite{aerts2009a}. We recall Theorem 4 of \cite{aerts2009a}.

\begin{theorem}
The membership weights $\mu (A),\mu (B) $ and $\mu (A\ \mathrm{or}\ B)$ of
an item $X$ with respect to concepts $A$ and $B$ and their disjunction `$A$
or $B$' are classical disjunction data if and only if they satisfy the
following `classical disjunction' inequalities: 
\begin{eqnarray}
&&0\leq \mu (A)\leq \mu (A\ \mathrm{or}\ B)\leq 1  \label{disjunctionineq01}
\\
&&0\leq \mu (B)\leq \mu (A\ \mathrm{or}\ B)\leq 1  \label{disjunctionineq02}
\\
&&0\leq \mu (A)+\mu (B)-\mu (A\ \mathrm{or}\ B)  \label{disjunctionineq03}
\end{eqnarray}
\end{theorem}
Proof: See \cite{aerts2009a}.

\bigskip
\noindent
In the case of two concepts $A_{1}$, $A_{2}$ and their disjunction `$A_{1}\ 
\mathrm{or}\ A_{2}$' the set of indices is $S=\{(1,2)\}$ and the classical
disjunction polytope $d_{c}\left( n,S\right) $ is contained in the $2+|S|=3$
dimensional Euclidean space, i.e. $R(2,\{1,2\})={\mathbb{R}}^{3}$. Furthermore
we have four vectors $\epsilon \in \{0,1\}^{n}$, namely $(0,0),(0,1),(1,0)$
and $(1,1)$, and hence the four vectors $\overrightarrow{v}^{\epsilon }\in {%
\mathbb{R}}^{3}$ which are the following 
\begin{equation}
\overrightarrow{v}^{(0,0)}=(0,0,0)\quad \overrightarrow{v}%
^{(1,0)}=(1,0,1)\quad \overrightarrow{v}^{(0,1)}=(0,1,1)\quad 
\overrightarrow{v}^{(1,1)}=(1,1,1)
\end{equation}%
This means that the correlation polytope $d_{c}\left( n,S\right) $ is the
convex region spanned by the convex combinations of the vectors $%
(0,0,0),(1,0,1),(0,1,1)$ and $(1,1,1)$, and the disjunction vector is given
by $\overrightarrow{p}=(\mu (A_{1}),\mu (A_{2}),\mu (A_{1}\ \mathrm{or}\
A_{2}))$. It is well-known that every polytope admits two dual descriptions:
one in terms of convex combinations of its vertices, and one in terms of the
inequalities that define its boundaries. Following \cite{aerts2009a}, the inequalities
defining the boundaries for the polytope $d_{c}\left( 2,\{(1,2)\}\right) $
are given by:%
\begin{eqnarray}
&&0\leq p_{1}\leq p_{1\vee 2}\leq 1  \label{P1} \\
&&0\leq p_{2}\leq p_{1\vee 2}\leq 1  \label{P2} \\
&&0\leq p_{1}+p_{2}-p_{1\vee 2}\leq 1  \label{P3}
\end{eqnarray}%
We observe that the last inequality $p_{1}+p_{2}-p_{1\vee 2}\leq 1$ follows
easily because from $p_{1}\leq p_{1\vee 2}$ and $p_{2}\leq p_{1\vee 2}$
follows that $p_{1}+p_{2}-p_{1\vee 2}\leq p_{1\vee 2}\leq 1$ (again because
of (\ref{P1})).

\begin{theorem}
The classical disjunction inequalities formulated in theorem 4.1. are satisfied if and only if $%
\overrightarrow{p}\in d_{c}\left( 2,\{(1,2)\}\right) .$
\end{theorem}
Proof: Let us notice that%
\begin{eqnarray*}
\overrightarrow{p} &=&\left( 
\begin{array}{c}
p_{1} \\ 
p_{2} \\ 
p_{1\vee 2}%
\end{array}%
\right) =\left( 1-p_{1\vee 2}\right) \left( 
\begin{array}{c}
0 \\ 
0 \\ 
0%
\end{array}%
\right) +\left( p_{1\vee 2}-p_{1}\right) \left( 
\begin{array}{c}
0 \\ 
1 \\ 
1%
\end{array}%
\right) +\left( p_{1\vee 2}-p_{2}\right) \left( 
\begin{array}{c}
1 \\ 
0 \\ 
1%
\end{array}%
\right) +\left( p_{1}+p_{2}-p_{1\vee 2}\right) \left( 
\begin{array}{c}
1 \\ 
1 \\ 
1%
\end{array}%
\right)  \\
&=&a\left( 
\begin{array}{c}
0 \\ 
0 \\ 
0%
\end{array}%
\right) +b\left( 
\begin{array}{c}
0 \\ 
1 \\ 
1%
\end{array}%
\right) +c\left( 
\begin{array}{c}
1 \\ 
0 \\ 
1%
\end{array}%
\right) +d\left( 
\begin{array}{c}
1 \\ 
1 \\ 
1%
\end{array}%
\right) 
\end{eqnarray*}%
Hence if the classical disjunction inequalities formulated in theorem 4.1 are satisfied, then it is
easy to check that $\overrightarrow{p}\in d_{c}\left( 2,\{(1,2)\}\right) .$%
Vice versa, let $\overrightarrow{p}\in d_{c}\left( 2,\{(1,2)\}\right) .$
Rewriting $\overrightarrow{p}$ as above, and putting condition $0\leq
a,b,c,d\leq 1$ immediately follow the classical disjunction inequalities of theorem 4.1: 
\begin{equation*}
\begin{array}{c}
0\leq 1-p_{1\vee 2}\leq 1 \\ 
0\leq p_{1\vee 2}-p_{1}\leq 1 \\ 
0\leq p_{1\vee 2}-p_{2}\leq 1 \\ 
0\leq p_{1}+p_{2}-p_{1\vee 2}\leq 1%
\end{array}%
\end{equation*}%
The last inequality \textit{is} condition (\ref{P3}), while $0\leq 1-p_{1\vee
2}\Rightarrow p_{1\vee 2}\leq 1.$ Also $0\leq p_{1\vee 2}-p_{1}\Rightarrow
p_{1}\leq p_{1\vee 2}.$ Also, $0\leq p_{1}+p_{2}-p_{1\vee 2}$ implies that $%
p_{1\vee 2}-p_{2}\leq p_{1}$ and since $0\leq p_{1\vee 2}-p_{2}$ follows
that $0\leq p_{1\vee 2}-p_{2}\leq p_{1}$ so $0\leq p_{1}.$ Putting these
together, we obtain then $0\leq p_{1}\leq p_{1\vee 2}\leq 1.$ Similarly, we
can prove $0\leq p_{2}\leq p_{1\vee 2}\leq 1.$ {\bf QED}.

\bigskip
\noindent
Let us consider the experimental data in Table 1. In the first part of Figure 1 we have represented the disjunction vectors formed by the membership weights of the different items to be found in Table 1 with respect to the pairs of concepts {\it Home Furnishings} and {\it Furniture} and their disjunction `{\it Home Furnishings or Furniture}', and also the disjunction polytope. The classical items, hence with disjunction vector inside the polytope, are represented by a little open disk. They are {\it Desk}, {\it Bed}, {\it Rug}, {\it Wall-Hangings}, {\it Shelves}, {\it Sculpture}, {\it Bath Tub}, {\it Door Bell} and {\it Desk Chair}. The non classical items, hence with disjunction vector outside of the polytope, are represented by a little closed disk. They are {\it Lamp}, {\it Wall Mirror}, {\it Window Seat}, {\it Painting}, {\it Light Fixture}, {\it Mantelpiece}, {\it Refrigerator}, {\it Space Rack}, {\it Sink Unit}, {\it Waste Paper Basket}, {\it Kitchen Count}, {\it Bar}, {\it Hammock}, {\it Ashtray} and {\it Park Bench}.

In the second part of Figure 1 we have represented the disjunction vectors formed by the membership weights of the different items found in Table 1 with respect to the pairs of concepts {\it Spices} and {\it Herbs} and their disjunction `{\it Spices or Herbs}', and the disjunction polytope. The classical items, hence with disjunction vector inside the polytope, are again represented by a little open disk. They are {\it Parsley}, {\it Mint}, {\it Pepper}, {\it Cinnamon}, {\it Peppermint}, {\it Sesame Seeds}, {\it Poppyseeds}, {\it Molasses} and {\it Chocolate}. The non classical items, hence with disjunction vector outside of the polytope, are again represented by a little closed disk. They are {\it Chires}, {\it Oregano}, {\it Chili Pepper}, {\it Mustard}, {\it Horseradish}, {\it Turmeric}, {\it Root Ginger}, {\it Salt}, {\it Curry}, {\it MSG}, {\it Vinegar}, {\it Vanilla}, {\it Lemon Juice}, {\it Sugar} and {\it Saccharin}.

In the first part of Figure 2 we have represented the disjunction vectors formed by the membership weights of the different items to be found in Table 1 with respect to the pairs of concepts {\it Hobbies} and {\it Games} and their disjunction `{\it Hobbies or Games}', and also the disjunction polytope. The classical items, hence with disjunction vector inside the polytope, are represented by a little open disk. They are {\it Monopoly}, {\it Tennis}, {\it Bridge}, {\it Bowling}, {\it Fishing}, {\it Theatre Going}, {\it Gardening} and {\it Guitar Playing}. The non classical items, hence with disjunction vector outside of the polytope, are represented by a little closed disk. They are {\it Noughts}, {\it Archery}, {\it Skating}, {\it Karate}, {\it Judo}, {\it Keep Fit}, {\it Jogging}, {\it Autograph Hunting}, {\it Discuss Throwing}, {\it Rock Climbing}, {\it Camping},  {\it Stamp Collection}, {\it Wrestling}, {\it Beer Drinking}, {\it Washing Dishes} and {\it Eating Ice-Cream Cones}.

In the second part of Figure 2 we have represented the disjunction vectors formed by the membership weights of the different items found in Table 1 with respect to the pairs of concepts {\it Games} and {\it Instruments} and their disjunction `{\it Games or Instruments}', and the disjunction polytope. The classical items, hence with disjunction vector inside the polytope, are represented by a little open disk. They are {\it Pliers}, {\it Tin Opener}, {\it Paint Brush}, {\it Meat Thermometer}, {\it Tuning Fork}, {\it Stapler}, {\it Magnetic Compass} and {\it Toothbrush}. The non classical items, hence with disjunction vector outside of the polytope, are represented by a little closed disk. They are {\it Pen-Knife}, {\it Scalpel}, {\it Scissors}, {\it Bicycle Pump}, {\it Step Ladder}, {\it Spoon}, {\it Door Key}, {\it Measuring Calipers}, {\it Paper Clip}, {\it Computer}, {\it Pencil Eraser}, {\it Broom}, {\it Sellotape}, {\it Rubber Band}, {\it Goggles} and {\it Skate Board}.

In the first part of Figure 3 we have represented the disjunction vectors formed by the membership weights of the different items to be found in Table 1 with respect to the pairs of concepts {\it Pets} and {\it Farmyard Animals} and their disjunction `{\it Pets or Farmyard Animals}', and also the disjunction polytope. The classical items, hence with disjunction vector inside the polytope, are represented by a little open disk. They are {\it Colie Dog}, {\it Race Horse}, {\it Spider}, {\it Robin} and {\it Blue Tit}. The non classical items, hence with disjunction vector outside of the polytope, are represented by a little closed disk. They are {\it Chicken}, {\it Ginger Tom-Cat}, {\it Prize Bull}, {\it Cart Horse}, {\it Donkey}, {\it Doberman Guard Dog}, {\it Siamese Cat}, {\it Goldfish}, {\it Rat}, {\it Guide Dog fro the Blind}, {\it Homing Pig}, {\it Field Mouse}, {\it Fox}, {\it Husky in Sledream}, {\it Badger}, {\it Circus Horse}, {\it Monkey}, {\it Squirrel} and {\it Camel}.

In the second part of Figure 3 we have represented the disjunction vectors formed by the membership weights of the different items found in Table 1 with respect to the pairs of concepts {\it Fruits} and {\it Vegetables} and their disjunction `{\it Fruits or Vegetables}', and the disjunction polytope. The classical items, hence with disjunction vector inside the polytope, are again represented by a little open disk. They are {\it Pumpkin}, {\it Tomato}, {\it Yam}, {\it Green Pepper}, {\it Apple}, {\it Chili Pepper}, {\it Peanut} and {\it Black Pepper}. The non classical items, hence with disjunction vector outside of the polytope, are again represented by a little closed disk. They are {\it Broccoli}, {\it Mushroom}, {\it Watercress}, {\it Coconut}, {\it Raisin}, {\it Olive}, {\it Elderberry}, {\it Lentils}, {\it Root Ginger}, {\it Garlic}, {\it Parsley}, {\it Almond}, {\it Rice}, {\it Acorn}, {\it Wheat} and {\it Mustard}.

In the first part of Figure 4 we have represented the disjunction vectors formed by the membership weights of the different items to be found in Table 1 with respect to the pairs of concepts {\it Sportswear} and {\it Sports Equipment} and their disjunction `{\it Sportswear or Sports Equipment}', and also the disjunction polytope. The classical items, hence with disjunction vector inside the polytope, are represented by a little open disk. They are {\it Golf Ball}, {\it Tennis Racket}, {\it Baseball Bat}, {\it American Foot}, {\it Rubber Flipper}, {\it Running Shoes}, {\it Sailing Life Jacket}, {\it Backpack} and {\it Cheerleaders}. The non classical items, hence with disjunction vector outside of the polytope, are represented by a little closed disk. They are {\it Diving Mask}, {\it Wrist Sweat}, {\it Frisbee}, {\it Gymnasium}, {\it Lineman's Flag}, {\it Ballet Shoes}, {\it Motorcycle Helmet}, {\it Bathing Costume}, {\it Underwater}, {\it Hoola Hoop}, {\it Cricket Pitch}, {\it Referee's Whistle}, {\it Sunglasses}, {\it Circus Clowns} and {\it Suntan Lotion}.

In the second part of Figure 4 we have represented the disjunction vectors formed by the membership weights of the different items found in Table 1 with respect to the pairs of concepts {\it Household Appliances} and {\it Kitchen Utensils} and their disjunction `{\it Household Appliances or Kitchen Utensils}', and the disjunction polytope. The classical items, hence with disjunction vector inside the polytope, are again represented by a little open disk. They are {\it Rolling Pin}, {\it Carving Knife}, {\it Whisk}, {\it Food Processor}, {\it Blender}, {\it Toaster}, {\it Spatula}, {\it Cake Tin}, {\it Cooking Stove}, {\it Rubbish Bin}, {\it Vacuum Cleaner}, {\it Vegetable Rack}, {\it Broom} and {\it Apron}. The non classical

\begin{figure}[H]
\centerline {\includegraphics[width=16cm]{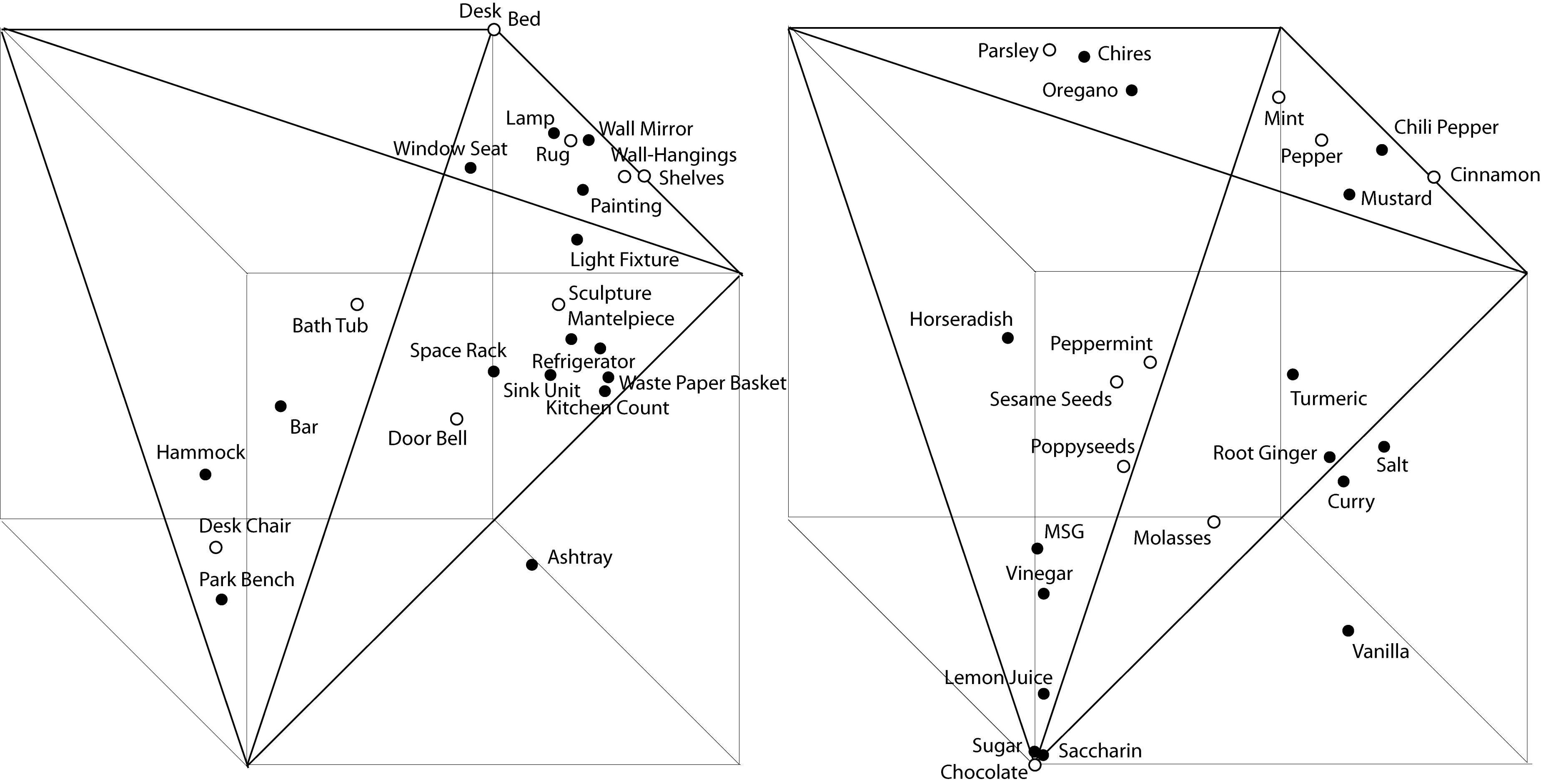}}
\caption{The polytopes for the concepts {\it Home Furnishing} and {\it Furniture} and the concepts {\it Spices} and {\it Herbs}. The classical items correspond to an open disk while the quantum ones to a full disk.}
\end{figure}

\begin{figure}[H]
\centerline {\includegraphics[width=16cm]{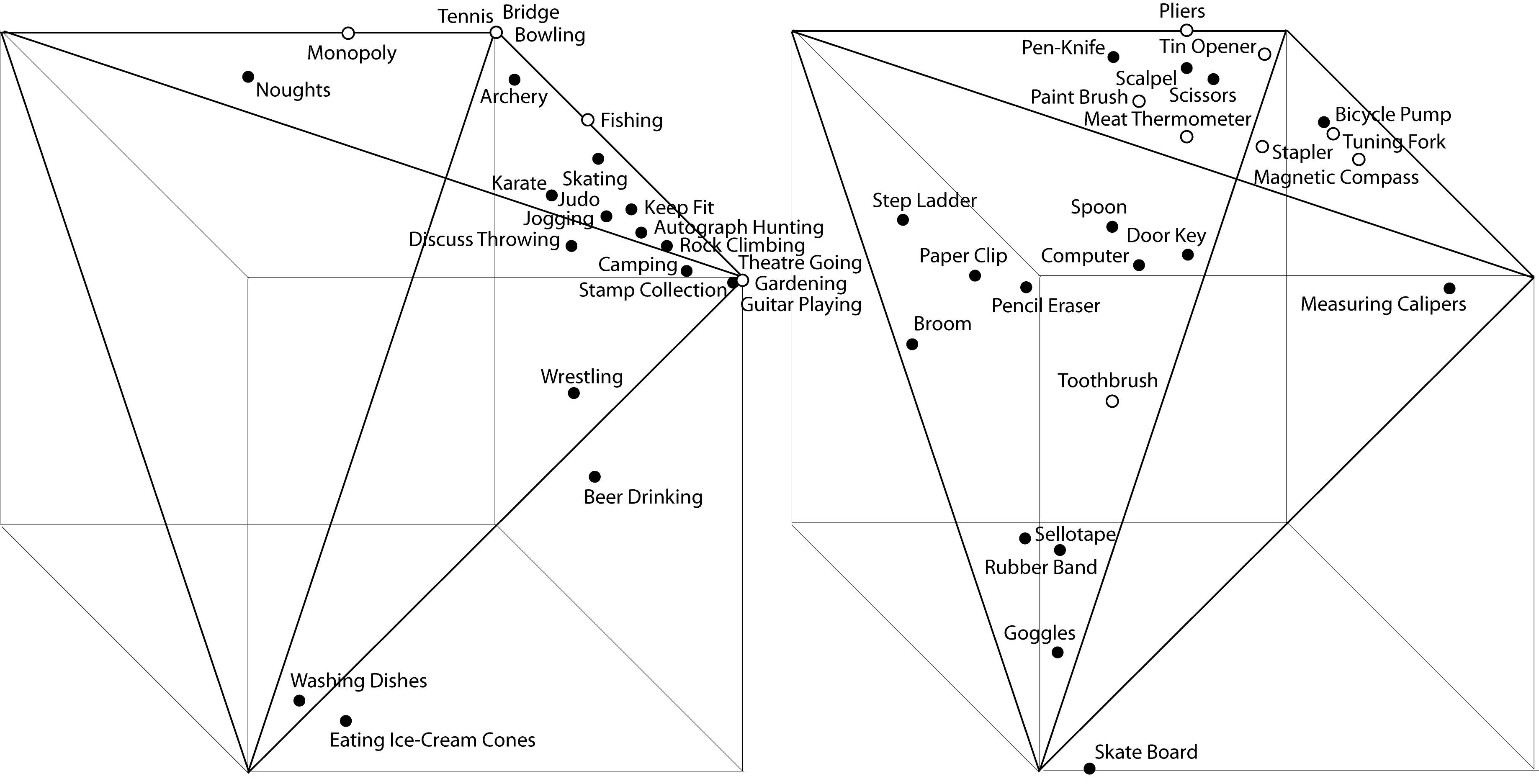}}
\caption{The polytopes for the concepts {\it Hobbies} and {\it Games} and the concepts {\it Instruments} and {\it Tools}. The classical items correspond to an open disk while the quantum ones to a full disk.}
\end{figure}

\begin{figure}[H]
\centerline {\includegraphics[width=16cm]{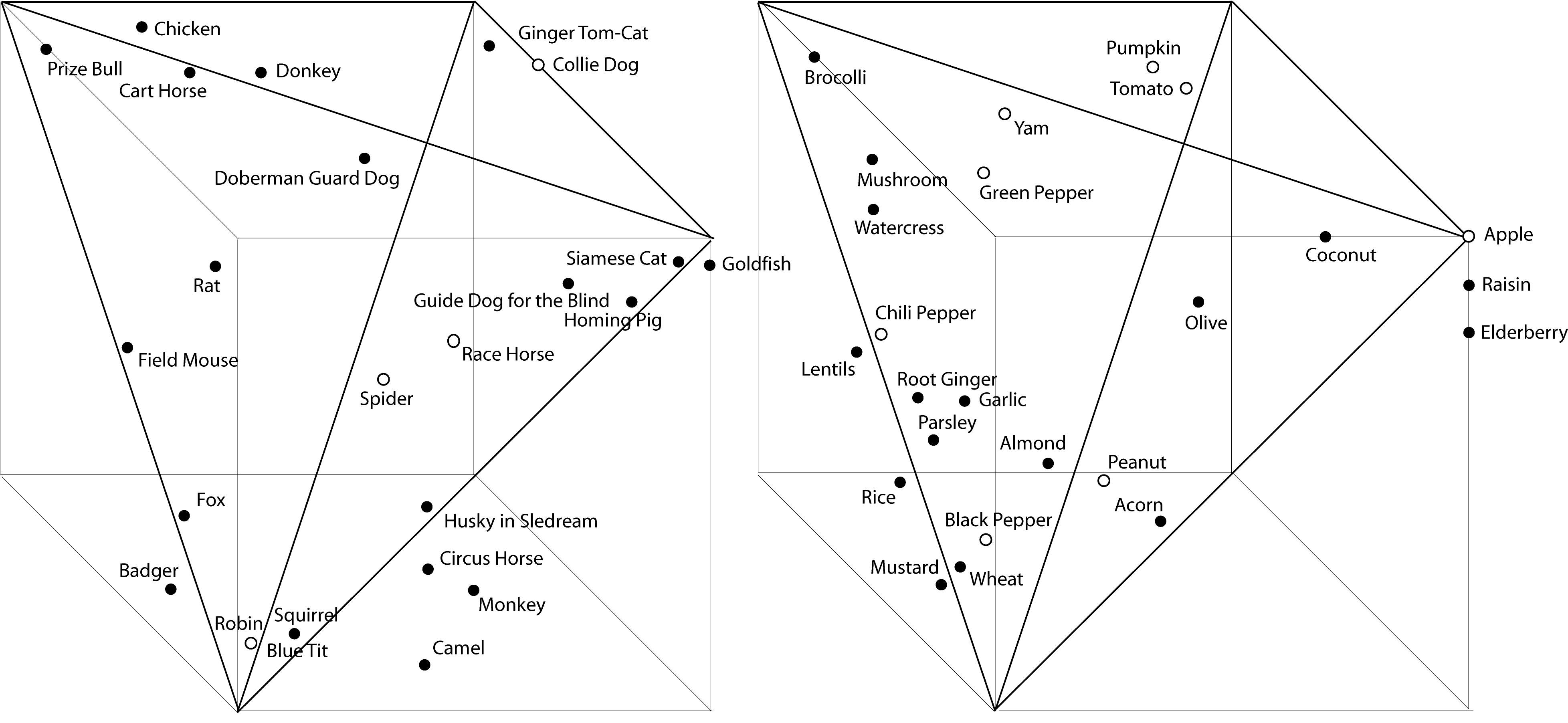}}
\caption{The polytopes for the concepts {\it Pets} and {\it Farmyard Animals} and the concepts {\it Fruits} and {\it Vegetables}. The classical items correspond to an open disk while the quantum ones to a full disk.}
\end{figure}

\begin{figure}[H]
\centerline {\includegraphics[width=16cm]{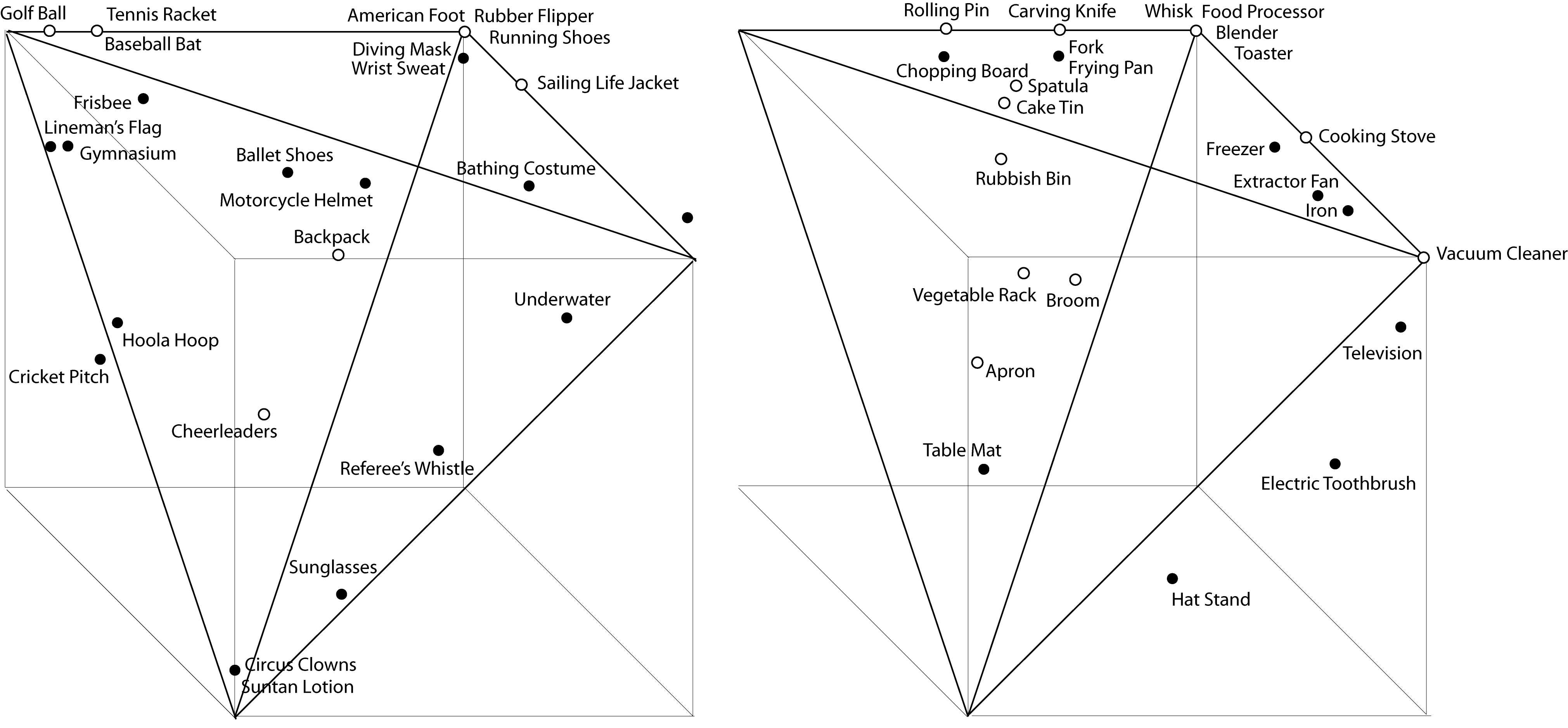}}
\caption{The polytopes for the concepts {\it Sportswear} and {\it Sports Equipment} and the concepts {\it Household Appliances} and {\it Kitchen Utensils}. The classical items correspond to an open disk while the quantum ones to a full disk.}
\end{figure}

\noindent
items, hence with disjunction vector outside of the polytope, are again represented by a little closed disk. They are {\it Fork}, {\it Frying Pan}, {\it Chopping Board}, {\it Freezer}, {\it Extractor Fan}, {\it Iron},  {\it Television}, {\it Electric Toothbrush}, {\it Table Mat} and {\it Hat Stand}.

Remark that if the experimental data turn out such that the item is a classical item, this does not mean that perhaps non classical effect are at play also for this item. But the non classical effect might be such that they do not show up with these particular measurements. This aspect of the situation is analyzed in more detail in \cite{aerts2009a}.

The inequalities that define the boundaries of polytope $d_{c}\left(
n,S\right) $ are a variant of the well-known Bell inequalities \cite%
{bell1964,pitowsky1989}, studied in the foundations of quantum mechanics, but now
put into the context of disjunctive connectives instead of conjunctive
correlations. This means that the violation of these inequalities, such as
it happens by the data corresponding to items for which the points lie
outside the polytope, has from a probabilistic perspective an analogous
meaning as the violation of Bell inequalities for the conjunction. Hence
these violations may indicate the presence of quantum structures in the
domain where the data is collected, which makes it plausible that a quantum
model, such as for example the one proposed in \cite{aerts2009a}, can be used
to model the data.

As we have shown above, the classical disjunction polytope allows for one
necessary and sufficient condition $\overrightarrow{p}\in d_{c}\left(
n,S\right) $ which guarantees a classical Kolmogorovian model for the given
set of probabilities to exist \cite{kolmogorov1956}. As illustrated here, this can be expressed by
a set of Bell-like inequalities. However, as Pitowsky remarked \cite{pitowsky1989}, the number
and complexity of the inequalities will grow so fast with $n$, that it would
require exponentially many computation steps to derive them all. Anyway,
already for the simplest (non-trivial) case $n=2$ interesting inequalities
can be derived by which the non classical nature of a set of statistical
data can be demonstrated explicitly. Such data exists in various fields of
science: of course in quantum mechanics, but also in cognition (concept)
theory, decision theory and some paradoxical situations in economics, such
as in the Allais and Ellsberg paradox situations \cite{allais1953,ellsberg1961}, notably situations which violate Savage's `Sure-Thing principle' \cite{savage1954}.

\bigskip
\noindent
{\bf Acknowledgments}

\noindent
This work was supported by grants G.0405.08 and G.0234.08 of the Research Program of the Research Foundation-Flanders (FWO, Belgium).

\end{document}